# TREATMENT OF THE QUASI-HARMONIC POTENTIAL WITH THE CENTRIFUGAL TYPE TERM IN THE SCHRÖDINGER EQUATION VIA LAPLACE TRANSFORM

D. R. CONSTANTIN[1], V. I. R. NICULESCU[2]

[1]Astronomical Institute of Romanian Academy, Bucharest, Romania,
E-mail: ghe12constantin@yahoo.com

[2] National Institute for Lasers, Plasma and Radiation Physics, Bucharest, Romania,
E-mail: filo_niculescu@yahoo.com

*Abstract.* In the quantic frame, for $D = 3$ dimensional space, in the two body problem case, we approach the Schrödinger equation (SE) taking in account the potential: $V_q(r) = \Delta r^2 + \frac{A}{r} + \frac{B}{r^2}$ called by us quasi-harmonic potential with the centrifugal type term $\frac{B}{r^2}, \Delta, A, B > 0$ and $\Delta \ll 1$. We use Laplace transform method (LTM) and we find for the first time an analytic solution of the $V_q$ potential problem. Namely, using directly and inverse Laplace transformations, we obtain the complete forms of the energy eigenvalues and wave functions. Furthermore, for this potential $V_q$, we make considerations about critical orbital quantum value $\ell_c$ and we obtain a useful approximation of upper bound $\ell_c^+$ to $\ell_c$.

*Key words:* Schrödinger equation; Laplace transform; analytic eigenfunctions; quasi harmonic potential with centrifugal type term

*PACS:* 03.65.Ge, 02.30.Uu, 02.30.Hq, 02.90.+p

## 1. INTRODUCTION

In the quantum mechanics, several authors solved Schrödinger equation [1] for some potentials like pseudoharmonic [2], Coulomb like [3], Mie type [4] and some nuclear potentials like Halton, Manning Rosen [5], Pöschel-Teller [6] and Woods-Saxon potentials [7-8].

There are a lot of methods for solving the Schrödinger equation such as Laplace transform method [9], Nikiforov-Uvarov method [10], homotopy perturbation method [11], series solution method [12], Fourier transform method [13], asymptotic iteration method [14], super-symmetric approach [15], variational method [16] and other methods.

The prior part in the present paper is to point out that Laplace transforms leads to the analytic and exact forms of eigenfunctions for the potential proposed by us:

$$V_q(r) = \Delta r^2 + \frac{A}{r} + \frac{B}{r^2}$$

which is a special quasi harmonic type potential containing a centrifugal type term. Choosing the coefficients $\Delta, A, B > 0$, $V_q$ potential implies a short range potential behavior. In the nuclear potential case, we mention there are a finite number of bound states beyond which the $\ell$ state is unbounded. From this reason, for a nuclear potential it's important to obtain the critical value of angular momentum $\ell_c$. So, the second aim of our paper is to obtain an approximation for upper bound range $\ell_c^+$ to $\ell_c$ for $V_q$ potential.

The present paper solves $V_q$ potential problem by Laplace transform method in the Section 2. In the Section 3, we communicate two categories of results: the energy and the eigenfunctions in the subsection 3.1., and also the computing of $\ell_c^+$ in the subsection 3.2. Conclusions close our paper in the Section 4.

## 2. BOUND STATES SPECTRUM VIA LAPLACE TRANSFORMATION METHOD

In the natural units $c = \hbar = 1$, assuming spherical symmetry of the potential, the *D* - dimensional time-independent Schrodinger equation in spherical coordinates $(r, \theta, \varphi)$ for a particle of mass *μ*, with arbitrary angular quantum number $\ell$, is given by [17, 10]:

$$\left[ -\frac{1}{2}\nabla^2 + V(r) \right] \Psi_{n\ell m}(r, \theta, \varphi) = E\Psi_{n\ell m}(r, \theta, \varphi) \qquad (1)$$

where *E* and *V(r)* denote the energy eigenvalues and potential.
The $\Psi_{n\ell m}(r, \theta, \varphi)$ denotes n-th state eigenfunctions.
The $\nabla^2$ in spherical coordinates is:

$$\nabla^2 = \frac{1}{r^2}\frac{\partial}{\partial r}\left(r^2\frac{\partial}{\partial r}\right) + \frac{1}{r^2\sin\theta}\frac{\partial}{\partial\theta}\left(\sin\theta\frac{\partial}{\partial\theta}\right) + \frac{1}{r^2\sin^2\theta}\frac{\partial^2}{\partial\varphi^2} \qquad (2)$$

We choose the bound state eigenfunctions $\Psi_{n\ell m}(r,\theta,\varphi)$ such that wave functions are vanishing for $r\to 0$, $r\to\infty$. We look for separable solution of SE:

$$\Psi_{n\ell m}(r,\theta,\varphi) = \Re_{n\ell}(r) Y_\ell^m(\theta,\varphi) \qquad (3)$$

where $\Re_{n\ell}(r)$ is the radial function and $Y_\ell^m(\theta,\varphi)$ is the angular function.

The equation (2) provides two separated equations: the one is known as the hyperspherical harmonics equation and the other one the hyperradial or in short the "radial" equation which is:

$$\Re''(r) + \frac{2}{r}\Re'(r) + \left[-\frac{\ell(\ell+D-2)}{r^2} + 2\mu(E-V(r))\right]\Re(r) = 0 \qquad (4)$$

where $\ell(\ell+D-2)|_{D=3}$ is the separation constant with $\ell = 0,1,2,3,...$ [18].

Taking in account the $V_q$ quasi-harmonic potential and $D=3$, the eq. (2) becomes:

$$\Re''(r) + \frac{2}{r}\Re'(r) + \left[-\frac{\ell(\ell+1)}{r^2} + 2\mu(E-\Delta r^2 - \frac{A}{r} - \frac{B}{r^2})\right]\Re(r) = 0 \qquad (5)$$

Taking $r\to\infty$, the formula (5) gets an asymptotic form:

$$\Re''(r) - 4d^2 r^2 \Re(r) = 0, \ \Delta>0, \ d = \sqrt{\frac{\mu\Delta}{2}} \qquad (6)$$

where the type term $4d^2$ is this chosen from mathematical conveniences.

We propose to find the solution by type:

$$\Re(r) = r^k e^{-dr^2} f(r), \quad k > 0 \qquad (7)$$

Hence, the function $f(r)$ and $k$-value remain to be determined.

Inserting the solution form (7) into the eq. (5), we have:

$$rf''(r) + \left(\eta_k + 2r - 2dr^2 - dr^3\right)f'(r) + \\ + \left[\frac{Q_{n\ell}}{r} - 2\mu A + \in_k r - dkr^2 - 2\mu\Delta r^3 + 2d^2 r^4\right]f(r) = 0 \quad (8)$$

the prime over $f(r)$ denoting the derivative with respect to $r$.
We introduce the following notations:

$$Q_{n\ell} = k(k+3) - \ell(\ell+1) - 2\mu B, \\ \in_k = 2\mu E - 2dk - 6d, \quad \eta_k = 2k \quad (9)$$

Starting from this point, we impose some parametric restrictions.
The first one is:

$$Q_{n\ell} = 0 \quad (10)$$

If we consider the following condition $\mu B = \ell$, then the eq.(10) implies two values for $k$:

$$k_+ = \ell \\ k_- = -(\ell + 3) \quad (11)$$

The acceptable physical value remains $k_+ = \ell$. Thus, the equation (8) becomes:

$$rf''(r) + \left(\eta_\ell + 2r - 2dr^2 - dr^3\right)f'(r) + \\ + \left[2d^2 r^4 - 2\mu\Delta r^3 - d\ell r^2 + \in_\ell r - 2\mu A\right]f(r) = 0 \quad (12)$$

We continue in the parametric restriction $d \ll 1$ and obtain for unknown function $f(r)$ the following differential equation:

$$rf''(r) + (\eta_\ell + 2r - 2dr^2)f'(r) + (-d\ell r^2 + \in_\ell r - 2\mu A)f(r) = 0 \quad (13)$$

We have to mention, we use the $Q_{n\ell} = 0$ and $d \ll 1$ conditions before applying Laplace transformation in the eq. (8), in order to reduce the differential equation from third to the second order in the transform space.

Therefore, we apply the Laplace transform [3] $\Phi(s) = L\{f(r)\}(s)$ where $\text{Re}(s) > 0$ and the eq. (13) becomes:

$$d(\ell - 2s)\Phi''(s) + (s^2 + 2s + \widetilde{\in})\Phi'(s) + (\gamma s + \alpha)\Phi(s) = 0 \quad (14)$$

where :
$$\gamma = 2 - 2\ell$$
$$\alpha = 2\mu A + 2$$
$$\widetilde{\in} = \in -4d . \quad (15)$$

We observe that $s_0 = \dfrac{\ell}{2}$ is the singular point.

This suggests the following form of the Laplace transform:

$$\Phi(s) = \frac{C_{n\ell}}{\left(s - \dfrac{\ell}{2}\right)^{n+1}} \quad (16)$$

And the inverse Laplace transform $f(r) = L^{-1}\{\Phi(s)\}(r)$ leads to the following expression:

$$f(r) = \frac{C_{n\ell}}{n!} r^n e^{\frac{\ell r}{2}} \quad (17)$$

## 3. RESULTS

### 3.1 ENERGY EIGENVALUES AND WAVE FUNCTIONS

Using the form (16) in the eq.(14), we obtain the system of conditions:

$$\gamma = n+1$$
$$4(n+1) + \ell\gamma = 2\alpha$$
$$\widetilde{N} + 2(n+1)\widetilde{\in} + \alpha\ell = 0, \tag{18}$$

where $\widetilde{N} = 4d(n+1)(n+2)$.

Solving the third equation from the system (18), we obtain the energy eigenvalues:

$$E_{n\ell} = \frac{d}{\mu}\left[\ell - n + 3 - \frac{\alpha\ell}{4d(n+1)}\right] = \sqrt{\frac{\Delta}{2\mu}}\left[\ell + 3 - n - \ell\frac{(4+\ell)}{8d}\right] \tag{19}$$

We remark that in the case of harmonic oscillator $V = \frac{1}{2}\mu\omega^2 r^2$, the energy expression (19) leads to the eigenvalue $E_{00} = \frac{3}{2}\omega$ which is the well known ground state energy.

A complete solution of SE implies the computation of normalization constant $C_{n\ell}$ from the condition:

$$\int_0^\infty [\Re(r)]^2 r^2 dr = 1 \tag{20}$$

Considering the approximation $r - \frac{\ell}{2d} \sim r$, we can evaluate the condition (20) by the integral formula: [3]

$$\int_0^\infty x^p e^{-ax^q} dx = \frac{1}{q}\frac{\Gamma\left(\frac{p+1}{q}\right)}{a^{\frac{p+1}{q}}}, \quad p, q, a > 0 \tag{21}$$

where $\Gamma(\cdot)$ is the gamma function. Thus, we obtain the normalization constant:

$$C_{n\ell} = n! \left\{ \frac{2(2d)^{\ell+n+\frac{3}{2}}}{\Gamma(\ell+n+\frac{3}{2})} e^{-\frac{\ell^2}{8d}} \right\}^{\frac{1}{2}} \tag{22}$$

Finally, using the relations (11), (17), (22) into the radial function (7), we find the complete analytic form of the eigenfunctions of the system :

$$\Re_{n\ell}(r) = r^{n+\ell} \frac{C_{n\ell}}{n!} e^{-dr^2+\frac{l}{2}r} \tag{23}$$

Further, we compute the $r_{rms}$ radius:

$$r_{rms} = \sqrt{<r^2>} \tag{24}$$

where

$$<r^2> = \int_0^\infty r^2 [\Re_{n\ell}(r)]^2 r^2 dr. \tag{25}$$

and obtain :

$$r_{rms} = \sqrt{\frac{\ell+n+\frac{3}{2}}{2d}} = \sqrt{\frac{\ell+n+\frac{3}{2}}{\sqrt{2\mu\Delta}}}, \Delta \neq 0 \tag{26}$$

### 3.2 UPPER BOUND APROXIMATION

The methodology for the SE using LT method, it is also useful in the short range potential case [19].
Taking in account $V_q$ as a nuclear potential (with $\Delta, A, B > 0$ ) and considering into the eq. (4) the transformation $\Re(r) = \frac{U(r)}{r}$, in the unit $2\mu = 1$, we are left with: [20]

$$\left[-\frac{d^2}{dr^2} - \lambda V_q(r) + \frac{\ell(\ell+1)}{r^2}\right] U_{n\ell}(r) = \widetilde{E}_{n\ell} U_{n\ell}(r) \tag{27}$$

Combining $\lambda > 0$ and $V_q$ (build that way to be an attractive potential), they give the type term $-\lambda V_q$ which is the short range potential; $\lambda$ is the strength potential. We consider the effective potential: [21]

$$V_{eff}(r) = -\lambda V_q(r) + \frac{\ell(\ell+1)}{r^2} \qquad (28)$$

A possible bound state of positive energy corresponds to the proposed $V_q$ potential including quasi bound states where $\lambda$ is relevant in the binding of these $\ell$-states. An infinitely small but negative part of $V_{eff}$ would admit a bound state. That means the critical strength $\lambda_c(\ell)$ is needed to bind a $\ell$ state.
Regarding of the above, we make some computations such as :
The radius r'$_0$, such that $V''_{eff}(r'_0) = 0$, leads to:

$$r'_0 = -3\frac{V'_q(r'_0)}{V''_q(r'_0)} = \frac{1}{2}\sqrt[3]{\frac{A}{\Delta}} > 0 \qquad (29)$$

The critical straight $\lambda_c$, which is the minimal value necessary to get a bound state, satisfies the condition:

$$\lambda_c \geq \ell(\ell+1)\frac{2}{-r_0^{'3}V'_q(r'_0)}, V_q(r'_0) \neq 0 \qquad (30)$$

We obtain the limit of large upper bound $\ell_c^+$ as:

$$\ell_c^+ \approx \sqrt{\frac{-r_0^{'3}V'_q(r'_0)}{2}}\sqrt{\lambda} = \sqrt{\frac{1}{2}\left(2B + \frac{3}{4}r'_0 A\right)}\sqrt{\lambda}, \qquad (31)$$

this upper bound approximation being suitable in the practical work.

## 4. CONCLUSIONS

In the two body problem associated to $V_q$ quasi-harmonic potential with $0 < \Delta \ll 1$, after some parametric restrictions, we solved 3-dimensional Schrödinger equation via Laplace transformation method. We obtained a complete analytic solution, namely the energy eigenvalues and eigenfunctions.

In the computation of $k$ positive value, we need to consider the condition $\mu B = \ell$ and this leaded to $k_+ = \ell$. Otherwise, the $k$ value, involved in parametric restriction $Q_{n\ell} = 0$, may be a subject of interest in quantum computing.

Using the wave function, we computed the root-mean-square (*rms*) charge radius which is measured for most stable nuclei by electron scattering form factors and/or from the x-ray transition energies of muonic atoms. The formula of $r_{rms}$ obtained by us is interesting to be compared with the experimental results.

Considering $V_q$ potential as a nuclear potential ($\Delta, A, B > 0$), we also obtained the analytic solution of the Schrödinger equation. This solution is only valid for well bound states but not for angular momentum close to or above the critical $\ell_c$. We estimated also the $\lambda_c(\ell)$ representing the necessary strength of the $V_q$ potential required such that $V_{eff}$ to have a negative part. Therefore, we calculated an upper bound $\ell_c^+$ which is a good approximation of $\ell_c$ and we find $\ell_c^+$ it is proportional to $\sqrt{\lambda}$ strength of potential.

*Acknowledgment:* Special thanks to Agneta Mocanu for helpful and fruitful discussions. This work was supported by a grant of the Romanian National Authority for Scientific Research and Innovation, CNCS/CCCDI - UEFISCDI, project number PN-III-P2-2.1-PED-2016-1189, within PNCDI III and also, by a grant of the Ministry of National Education and Scientific Research, RDI Programe for Space Technology and Avanced Research - STAR, project number 513.

## REFERENCES

[1] E. Schrodinger, *An undulatory theory of the mechanics of atoms and molecules,* Ann.der Physik. **28**, 6, 1049-1070 (1926)

[2] Al. Arda, R. Sever, *Exact Solutions of the Schrodinger Equation via Laplace Transform Approach: Pseudoharmonic potential and Mie-type potentials*, http://arxiv.org/abs/1202.4268v1, 1-15 (2012)

[3] M. J. Englefield, *Solution of Coulomb Problem by Laplace Transform,* Journal of Mathematical Analysis and Applications **48**, 210-275 (1974)

[4] T. Das, *Treatment of N-dimensional Schrödinger Equation for Anharmonic Potential via Laplace Transform*, EJTP **13**, 35, 207–214 (2016)

[5] A. N. Ikot1, L. E. Akpabio, E. B. Umoren, *Exact Solution of Schrödinger Equation with Inverted Woods-Saxon and Manning-Rosen Potential*, J. Sci. Res. **3 (1)**, 25-33 (2011)

[6] H.Kleinert, I. Mustapic, *Summing the spectral representations of Poschl-Teller and Rosen-Morse fixed-energy amplitudes*, J.Math. Phys, **33**(2), 643-661 (1992)

[7] M.C, Apak, D Petrellis, B.Gonül1, D. Bonatsos, *Analytical solutions for the Bohr Hamiltonian with the Woods–Saxon potential*, http://arxiv.org/abs/1506.07444v1, 1-24 (2015)

[8] M. Mirea, *Microscopic treatments of fission inertia within the Woods-Saxon Two center shell model*, Romanian Reports in Physics, **63**, 3, 676-684 (2011)

[9] S. Murray R. Spiegel, *Laplace Transforms*, Rensselaer Polytechnic Institute (1965)

[10] Cüneyt Berkdemir, Application of the Nikiforov-Uvarov Method in Quantum Mechanics, Theoretical Concepts of Quantum Mechanics, InTech, 226-251 (2012)

[11] S. Al-Jaber, *Solution of the radial n-dimensional Schrödinger equation using homotopy perturbation method*, Rom. Journ. Phys., Vol. **58**, Nos. 3–4, 247–259 (2013)

[12] Viorica Florescu, Quantum Mechanics Lectures I, Editura Universităţii, Bucharest, 2007, (in Romanian).

[13] M. Rizea, *Fourier transforms of single-particle wave functions in cylindrical coordinates*, (2016)

[14] M. Aygun, O. Bayrak, I. Boztosun, Solution of the Radial Schrödinger Equation for the $V_k$ Potential Family using the Asymptotic Iteration Method, http://arxiv.org/abs/math-ph/0703040v1, 1-12 (2007)

[15] F. Cooper, A. Khare ,U. Sukhatme, *Supersymmetry and Quantum Mechanics,* Phys.Rep., 251-267 (1995)

[16] B. H. Bransden, C. J. Joachain, Introduction in Quantum Mechanics, Editura Tehnica, Bucharest, 1995, p 95 (in Romanian)

[17] S. H. Dong, *Wave Equations in Higher Dimensions* Springer, New York (2011)

[18] N. Shimakura, *Partial Differential Operator of Elliptic Type*, American Mathematical Society (1992)

[19] V. I. R. Niculescu, D. Popescu, R. Anton, L. Sandru, A *new family of Woods-Saxon potentials with complex poles,* Rom. Journ. Phys., **61**, 1513–1518 (2016)

[20] A. Diaf, M. Lassaut, R.J. Lombard, *Approximate centrifugal barriers and critical angular momentum*, Rom. Journ. Phys., Vol. **57,** 1-2, 159-164 (2012)

[21] F. Calogero, *Sufficient conditions for an attractive potential to possess bound states*, Journal of Mathematical Physics, **6**, 1, 161-164 (1965)